\documentclass[aps,prd,preprintnumbers,nofootinbib,floatfix,11pt]{revtex4}

\usepackage{amsfonts,amsmath,amssymb,mathrsfs,graphicx,epsfig,color,amsthm}

\allowdisplaybreaks[1]
\newcommand{\be}{\begin{equation}}
\newcommand{\ee}{\end{equation}}
\newcommand{\ba}{\begin{eqnarray}}
\newcommand{\ea}{\end{eqnarray}}
\newcommand{\nn}{\nonumber}

\theoremstyle{definition}

\usepackage{}

\usepackage[normalem]{ulem}

\newcommand{\red}[1]{{\textcolor{red}{#1}}} 
\newcommand{\blue}[1]{{\textcolor{blue}{#1}}}

\newcommand{\Lag}{{\cal L}}
\newcommand{\re}{{\mbox{Re\,}}}

\newcommand{\tr}{{\mbox{Tr\,}}}
\newcommand{\scl}{{\sigma_{\rm cl}}}

\begin{document}

\title{
UV stability of one-loop radiative corrections\\
in higher-derivative scalar field theory
}

\author{Yugo Abe${}^1$, Takeo Inami${}^{2,3}$, Keisuke Izumi${}^{4,5}$}
\affiliation{$^{1}$Department of Pure and Applied Physics, Kansai University, Suita 564-8680, Japan}
\affiliation{$^{2}$iTHEMS Program, RIKEN, Wako 351-0198, Japan}
\affiliation{$^{3}$Yukawa Institute for Theoretical Physics, Kyoto University,
Kitashirakawa Oiwakecho, Sakyo-ku, Kyoto 606-8502, Japan}
\affiliation{$^{4}$Department of Mathematics, Nagoya University, Nagoya 464-8602, Japan}
\affiliation{$^{5}$Kobayashi-Maskawa Institute, Nagoya University, Nagoya 464-8602, Japan} 

\begin{abstract}
\begin{center}
{\bf Abstract}
\end{center}
We consider the theory of a higher-derivative (HD) real scalar field $\phi$ coupled to a complex scalar $\sigma$, the coupling of the $\phi$ and $\sigma$ being given by two types, $\lambda_{\sigma\phi}\sigma^\dagger \sigma\phi^{2}$ and $\xi_{\sigma\phi}\sigma^\dagger \sigma\left(\partial_{\mu}\phi\right)^{2}$. 
We evaluate $\phi$ one-loop corrections $\delta V(\sigma)$ to the effective potential of $\sigma$, both the contribution from the positive norm part of $\phi$ and that from the {\it negative norm part} (ghost). 
We show that $\delta V(\scl)$ at $\sigma_{\rm cl}\to \infty$, where $\sigma_{\rm cl}$ is a classical value of $\sigma$, is positive, implying the stability of $\delta V(\scl)$ by the HD one-loop radiative corrections at high energy.
\end{abstract}
\maketitle


\section{Introduction}
Detailed analyses of the standard model (SM) incorporating higher loop corrections have suggested a possibility that the SM may be viable up to the Planck energy scale \cite{Hamada:2012bp,Bezrukov:2012sa,Degrassi:2012ry}. 
This possibility raises the question that gravitational interaction may matter in determining the Higgs potential through the graviton one loop \cite{Smolin:1979ca,Bhattacharjee:2012my,Haba:2014qca,Loebbert:2015eea,Abe:2016irv,Alvarez-Luna:2022hka}. 
This is particularly so when the SM Higgs potential (including radiative corrections) is small as implied by the analyses in Refs.~\cite{Hamada:2012bp,Bezrukov:2012sa,Degrassi:2012ry}. 
One needs a UV complete quantum gravity theory to evaluate graviton loop effects. 
One such theory is quadratic gravity by Stelle~\cite{Stelle:1976gc}. 
A review article by Salvio~\cite{Salvio:2018crh} gives a comprehensive literature of this topic. 

The quadratic gravity theory contains a massive spin-2 particle and a scalar, in addition to the canonical massless spin-2 graviton. 
The massive spin-2 field has a negative sign in the kinetic term%
\footnote{
The study of a Hamiltonian system with higher derivative kinetic term dates back to the work of Ostrogradsky in  mid-19th century \cite{Ostrogradsky:1850fid}.
Field theories with higher derivatives were first studied by Pais and  Uhlenbeck in Ref. \cite{Pais:1950za}. The related issues were recently reviewed in Ref. \cite{Woodard:2015zca}.
}
({\it i.e.}, it is a ghost field), which either leads to {\it infinite negative energy} or a {\it negative norm state}, due to its higher derivative (HD) nature, as reviewed in Ref. \cite{Salvio:2018crh}. We take the second view in this paper 
so that the theory becomes renormalizable and that quantization can be performed without meeting the question of negative energy. 
After studies of field theories possessing ghosts from a few different perspectives, initiated by Lee and Wick \cite{Lee:1969fy,Lee:1970iw},
the role of ghost yet lacks complete understanding, especially at the quantum level.
Lee and Wick discussed that the massive ghost is expected to decay into positive norm light particles, but the possibility of the existence of asymptotic ghost fields has recently been pointed out by Kubo and Kugo~\cite{Kubo:2023lpz,Kubo:2024ysu}.

Since the canonical massless graviton is a boson, its radiative corrections to the Higgs potential make its stability at high energy (we call it UV stability) better. 
However, ghost fields are expected to give opposite contributions, which is the origin of the cancelation of the loop divergence in HD theories. 
Then, it is not trivial that in HD theories, after the cancelation of the loop divergence, the remaining finite parts of the one-loop corrections are positive or not. 
Especially, the UV stability of the effective potential is required for the consistency of the theory, and thus it is critical to the UV completion of the quadratic gravity theory. 
Checking the UV stability in a HD theory is the main objective of this paper. 
To focus on the effect of a HD field,
we consider a simple model where a HD scalar field in interaction with an ordinary complex scalar field exist.

The theory of a HD scalar field $\phi$ \cite{Hawking:2001yt,Abe:2018rwb,Holdom:2023usn,Donoghue:2023yjt} is the simplest among field theories possessing the HD kinetic term. 
One can construct a model of a HD scalar field coupled to an ordinary complex scalar $\sigma$ (matter scalar) such that the model mimics the quadratic gravity coupled to the SM Higgs. 
In this paper we wish to study a specific problem, the quantum corrections from the HD scalar loop to the effective potential of the ``Higgs'' $\sigma$ \cite{Coleman:1973jx}.
It is interesting to see whether (and how) the effect of the HD loop is different from that of ordinary fields (of positive norm). 
Particularly, we focus on the UV stability of the HD-scalar one-loop correction to the effective potential of $\sigma$; it has not been studied in the past.

The paper is composed as follows. 
In the next section, we give a theory of a HD scalar field $\phi$ coupled to an ordinary complex scalar $\sigma$. 
Because of the HD kinetic term, a certain class of derivative couplings becomes renormalizable. 
Such derivative coupling term plays a role in the evolution of the one-loop correction to the effective potential $V(\sigma)$. 
Since the calculations to obtain the one-loop correction are involved, 
we show them in Appendix~\ref{App}. 
The final section is devoted to the summary and discussion.
The metric convention that we use is $(+1,-1,-1,-1)$.

\section{Model of a higher-derivative scalar field coupled to a complex scalar}

We consider a theory of a higher (fourth-) derivative (HD) scalar field $\phi$ in a four-dimensional spacetime ($d=4$) \cite{Hawking:2001yt,Abe:2018rwb,Holdom:2023usn,Donoghue:2023yjt} coupled to an ordinary {\it complex} (charged) scalar $\sigma$, thus introducing $U(1)$ symmetry. We construct a Lagrangian which is renormalizable in the sense of BPHZ by consulting the argument of Refs. \cite{Fujimori:2015mea,Abe:2018rwb},%
 \footnote{
Reference \cite{Fujimori:2015mea} deals with non-relativistic field theories and a slight change of the argument given there is needed. 
The extension to HD theories is shown in Ref. \cite{Abe:2018rwb}
 }
\begin{align}
\Lag&=\Lag_{\rm HD}(\phi)+\Lag_{\rm matter}(\sigma) + \Lag_{\phi{\mathchar`-}\sigma}(\phi,\sigma) .
\label{(2.1)}
\end{align}
Here $\Lag_{\rm HD}(\phi)$ is given by
\begin{align}
\Lag_{\rm HD}(\phi)=
-\frac{1}{2}\phi\left(\Box+m_{1}^{2}\right)\left(\Box+m_{2}^{2}\right)\phi-\frac{\lambda_{\phi}}{4!}\phi^{4}.
\label{(2.2)}
\end{align}
A wider class of self-interaction terms of $\phi$ than $\lambda_{\phi}\phi^{4}/4!$ is allowed from the renormalizability point of view \cite{Fujimori:2015mea,Abe:2018rwb}.
They will not play roles in our $\phi$ one-loop calculation and are not written in Eq.~\eqref{(2.2)}. $\Lag_{\rm matter}(\sigma)$ is given by 
\begin{align}
\Lag_{\rm matter}(\sigma) = \partial_\mu \sigma^\dagger  \partial^\mu \sigma - V(\sigma),
\label{(2.3)}
\end{align}
where
\begin{align}
V(\sigma)=m_{\sigma}^{2} \sigma^\dagger \sigma + \frac{\lambda_{\sigma}}{2}\left( \sigma^\dagger  \sigma \right)^{2}.
\label{(2.4)}
\end{align}
We restrict the $\phi{\mathchar`-}\sigma$ couplings to the renormalizable ones \cite{Abe:2018rwb},
\begin{align}
\Lag_{\phi{\mathchar`-}\sigma}(\phi,\sigma) = -\frac{\lambda_{\sigma\phi}}{2}(\sigma^\dagger \sigma)\phi^{2}-\frac{\xi_{\sigma\phi}}{2}(\sigma^\dagger \sigma)\left(\partial_{\mu}\phi\right)^{2} .
\label{(2.5)}
\end{align}
The dimensions of the coupling constants and the fields are
\begin{align}
[\lambda_{\sigma}]=0,\quad 
[\lambda_{\sigma\phi}]=2,\quad 
[\xi_{\sigma\phi}]=0;\quad
[\phi]=0,\quad 
[\sigma]=1.\label{(2.6)}
\end{align}
There are other types of $\phi{\mathchar`-}\sigma$ interaction of dimension 4, e.g. 
$(\sigma^\dagger \sigma)^2\phi^2$. However, they would generate divergent new terms by loop corrections and are not allowed from the renormalizability condition \cite{Abe:2018rwb,Fujimori:2015mea}. 
The vanishing of the dimension of $\phi$ ({\it i.e.} $[\phi]=0$ in Eq.~\eqref{(2.6)}) is due to the HD property of $\phi$ and is a key to the renormalizability of the theory with action~\eqref{(2.6)}. 
In Eq.~\eqref{(2.6)} $\lambda_{\sigma\phi}>0$ is required from the tree-level stability for $\sigma \to \pm \infty$, while the sign of $\xi_{\sigma\phi}$ may not be constrained.

The Lagrangian~\eqref{(2.2)} may be written in the second derivative form \cite{Hawking:2001yt,Abe:2018rwb}, 
\begin{align}
\Lag_{\rm HD}(\psi_1,\psi_2)=
-\frac{1}{2}\psi_{1}\left(\Box+m_{1}^{2}\right)\psi_{1}+\frac{1}{2}\psi_{2}\left(\Box+m_{2}^{2}\right)\psi_{2}-\frac{\lambda_{\phi}}{4! M^4}(\psi_2-\psi_1)^{4},
\label{(2.7)}
\end{align}
and
the interaction Lagrangian~\eqref{(2.5)} may become
\begin{align}
\Lag_{\phi{\mathchar`-}\sigma}(\phi,\sigma) = -\frac{\lambda_{\sigma\phi}}{2M^2}(\sigma^\dagger \sigma)(\psi_2-\psi_1)^{2}-\frac{\xi_{\sigma\phi}}{2M^2}(\sigma^\dagger \sigma)\left(\partial_{\mu}(\psi_2-\psi_1)\right)^{2} .
\label{psicoup}
\end{align}
Here, 
\begin{align}
\phi=\frac{\left(\psi_{2}-\psi_{1}\right)}{M}, \qquad
M:=\sqrt{m_{2}^{2}-m_{1}^{2}} .
\end{align}
Conversely, $\psi_{1}$ and $\psi_{2}$ are expressed in terms of $\phi$ as follows
\begin{align}
\psi_{1}=-\frac{\Box+m_{2}^{2}}{M}\phi~,\qquad\psi_{2}=-\frac{\Box+m_{1}^{2}}{M}\phi~.
\end{align}
The negative sign in $\psi_2$'s kinetic term
means that $\psi_2$ is a {\it negative norm} field.
In this form, the dimension of the $\psi_{1}$ and $\psi_{2}$ and that of the coupling constant for the derivative coupling are
\begin{align}
[\psi_{1}]=[\psi_{2}]=1,\quad[\xi_{\sigma\phi}/M^2]=-2.
\end{align}
In the sense of the power counting, the second term in the action (\ref{psicoup}) looks nonrenormalizable. 
However, due to the cancelation of the loop diagrams between $\psi_1$ and $\psi_2$, 
it turns out that this term is renormalizable~\cite{Abe:2018rwb}.


\section{one-loop corrections to the effective potential $V(\sigma)$ }

In HD theories, the cancelation of loop contributions occurs, and some derivative couplings become renormalizable. 
Then, natural questions arise.
How is the one-loop correction by HD fields?
What is the effect of the derivative couplings?
Is it different from the one-loop correction by the canonical fields?
Is the vacuum of the one-loop effective potential is stable?
To address these questions, we present a concrete calculation in the theory of complex scalar field $\sigma$ coupled to HD scalar field $\phi$,
which is described by the Lagrangian in Eq. \eqref{(2.1)}.
Here, the background fields $\sigma_{\rm cl}$ is assumed to be constant (with $\phi$ being assumed to have no V.E.V.).
Specifically, we calculate the $\phi$ one-loop correction to the effective potential of $\sigma$, 
denoted by $\delta V^{(\phi\, {\rm 1{\mathchar`-}loop})}(\sigma_{\rm cl})$. 
We will find that the UV stability of $\delta V^{(\phi\, {\rm 1{\mathchar`-}loop})}(\sigma_{\rm cl})$. 

\subsection{$\phi$ one-loop effective potential}

We consider the possibility that the ``Higgs'' $\sigma$ has a classical value $\scl$ whereas $\phi$ {\it does not}.
We simplify the notation by setting
\begin{align}
&\sigma \to \scl + \sigma \hspace{5mm} (\scl \mbox{ real}, \sigma \mbox{ complex})\label{(3.1)},\\
&\phi \to \phi  \hspace{14mm}(\mbox{{\it i.e.} $\phi_{\rm cl}=0$})\label{(3.2)}.
\end{align}
Here, we can set $\scl$ to have a real value.
Substitution of Eqs.~\eqref{(3.1)} and \eqref{(3.2)} into the Lagrangian~\eqref{(2.1)} gives us
\begin{align}
S=\int d^4x \Biggl\{
&-\frac{1}{2}\phi\left(\Box+m_{1}^{2}\right)\left(\Box+m_{2}^{2}\right)\phi
-  \sigma^\dagger  \left(\Box+m_{\sigma}^{2} + \lambda_{\sigma} \scl^2\right)\sigma
- 2 \left(m_\sigma^2 \scl + \lambda_\sigma \scl^3 \right) \left( \re \sigma \right)
\nn \\
&
-2 \lambda_\sigma \scl^2 \left(\re \sigma \right)^2 
-2\lambda_\sigma \scl \left( \re \sigma \right) \left( \sigma^\dagger  \sigma\right)
- \frac{\lambda_{\sigma\phi}}{2}\left[\scl^2 +2\scl \re\sigma +\left( \sigma^\dagger  \sigma\right) \right] \phi^2
\nn \\ 
&
- \frac{\xi_{\sigma\phi}}{2}\left[\scl^2 +2\scl \re\sigma +\left( \sigma^\dagger  \sigma\right) \right] \left(\partial_\mu \phi\right)^2
-\frac{\lambda_{\phi}}{4!}\phi^4 - \frac{\lambda_\sigma}{2}\left( \sigma^\dagger  \sigma\right)^2 -V^{(0)}\left(\scl\right)\Biggr\}.
\label{(3.3)}
\end{align}
The potential term $V^{(0)}\left(\scl\right)$, here, 
should be identified with the tree-level one, given by 
\begin{align}
V^{(0)}\left(\scl\right) =  m_\sigma^2 \scl^2 + \frac{\lambda_\sigma}{2} \scl^4.
\label{(3.4)}
\end{align}
After incorporating the one-loop correction, we have 
\begin{align}
V\left(\scl\right) = V^{(0)}\left(\scl\right) + \delta V^{(\phi\ {\rm 1{\mathchar`-}loop}) }\left(\scl\right).
\label{(3.5)}
\end{align}

We wish to address ourselves to the two questions: 
i) whether the potential~\eqref{(3.5)} is stable (bounded from below) or not and
ii) whether the $U(1)$ symmetry breaking will occur or not for Eq.~\eqref{(3.5)}, that is, $V'(\scl)=0$ and $V''(\scl)>0$ for a nonzero value of $\scl$ ($\scl\neq 0$).
\footnote{Since we have fixed the gauge degree of freedom associated with the $U(1)$ symmetry by choosing $\sigma_{\rm cl}$ to be real, the potential is expressed as a function of $\sigma_{\rm cl}^2$.
In a gauge-invariant notation, where the $U(1)$ symmetry remains manifest,
this should be understood as $|\sigma_{\rm cl}|^{2} \equiv \sigma_{\rm cl}^\dagger \sigma_{\rm cl}$.}
The second question can be answered after incorporating renormalization group effects at one-loop corrections \cite{Coleman:1973jx}.
This question will be addressed in the future work.

The $\phi$ one-loop correction $\delta V^{(\phi\ {\rm 1{\mathchar`-}loop}) }(\scl)$ of the potential of $\scl$ is obtained by the $\phi$-path integration of $\exp[iS^{(2)} (\phi)]$ (see, for instance, Ref. \cite{Laine:2016hma}), where $S^{(2)} (\phi)$ contains the terms with two powers of $\phi$ in action~\eqref{(3.3)}, and it is given by
\begin{align}
S^{(2)} (\phi) = \int d^4x \left(- \frac12 \phi(x) {\cal O}_\phi (x) \phi(x) \right),
\label{(3.6)}
\end{align}
where ${\cal O}_\phi (x)$ is the differential operator, 
\begin{align}
{\cal O}_\phi (x) := \Box{}^2 + \left( m_1^2 +m_2^2 - \xi_{\sigma\phi}\scl^2 \right) \Box+ m_1^2 m_2^2 + \lambda_{\sigma\phi} \scl^2. 
\label{(3.7)}
\end{align}
The $\phi$ one-loop correction is now given by 
\begin{align}
\delta V^{(\phi\ {\rm 1{\mathchar`-}loop}) }\left(\scl\right) = \frac12 \tr_{\phi} \ln \left\{\frac{{\cal O}_\phi (x)}{{\cal O}^{(0)}_\phi (x)}\right\}, 
\label{(3.8)}
\end{align}
where the free-field quadratic fluctuation operator for $\phi$, denoted by ${\cal O}^{(0)}_\phi (x)$, has been introduced. 
This addition is harmless because it is independent of $\scl$, and its role is simply to render the argument of the logarithm dimensionless in Eq.~\eqref{(3.8)}.
In the momentum space in Euclidean metric, the integral is written by replacing $\Box$ by $k_E^2$, that is,
\begin{align}
\Box \to -\Box_E =k_E^2, 
\end{align}
and then, 
we have  
\begin{align}
\delta V^{(\phi\ {\rm 1{\mathchar`-}loop}) }\left(\scl\right) 
=\frac12 \int  \frac{d^4 k_E}{(2\pi)^4} \ln \left[{\tilde {\cal O}}_{\tilde \phi} (k_E, \scl) \right]-\frac12 \int  \frac{d^4 k_E}{(2\pi)^4} \ln \left[{\tilde {\cal O}}_{\tilde \phi} (k_E, 0) \right],
\label{(3.9)}
\end{align}
where ${\tilde \phi} (k_E)$ is the Fourier basis of the $\phi (x)$ in Euclidean space and 
\begin{align}
{\tilde {\cal O}}_{\tilde \phi} (k_E, \scl) 
:= \left( k_E^2 \right)^2 + \left( m_1^2 +m_2^2 - \xi_{\sigma\phi}\scl^2 \right) k_E^2+ m_1^2 m_2^2 + \lambda_{\sigma\phi} \scl^2. 
\label{(3.10)}
\end{align}
Noting that $d^4 k_E=2 \pi^2 k_E^3 dk_E$, and after a change of variable $k_E^2=q$, we have 
\begin{align}
\frac12 \int  \frac{d^4 k_E}{(2\pi)^4} \ln \left[{\tilde {\cal O}}_{\tilde \phi} (k_E, \scl) \right] =
\frac1{32 \pi^2} \int_0^{\Lambda^2}  dq \, q \ln {\tilde {\cal O}} (q, \scl) ,
\label{(3.11)}
\end{align}
where we have set
\begin{align}
&{\tilde {\cal O}} (q, \scl):= [q+\alpha(\scl)]^2+ \beta(\scl),\label{(3.12)}\\
&\alpha(\scl):= \frac{m_1^2+m_2^2- \xi_{\sigma\phi}\scl^2}{2}, \label{(3.13)}  \\
&\beta(\scl) := - \frac{\left(m_2^2-m_1^2\right)^2}{4} + \left[ \lambda_{\sigma\phi} + \frac{\left(m_1^2+m_2^2\right) \xi_{\sigma\phi}}{2}\right]\scl^2 - \frac{\xi_{\sigma\phi}^2}{4} \scl^4,\nonumber\\
&~~~~~~~~\,=m_1^2 m_2^2 + \lambda_{\sigma\phi}\scl^2 - \left(\frac{m_1^2+m_2^2- \xi_{\sigma\phi}\scl^2}{2}\right)^2~.
\label{(3.14)}
\end{align}
The $k_E$ integration in Eq.~\eqref{(3.9)} is UV divergent. In this paper it is regularized by the momentum cutoff $\Lambda$.
\footnote{Other regularization are known to give the same result as the cutoff method. 
In Appendix~\ref{App of Dimensional Regularization}, we evaluate Eq.~\eqref{(3.9)} in the dimensional regularization, and we show that the finite terms coincide with the cutoff result.}

We now compute the logarithm integral~\eqref{(3.11)}. The integration can be done by setting $x=q+\alpha(\scl)$, and we obtain 
\begin{align}
\frac12 \int  \frac{d^4 k_E}{(2\pi)^4} \ln \left[{\tilde {\cal O}}_{\tilde \phi} (k_E, \scl) \right]
&= \frac{1}{32 \pi^2} \int _0^{\Lambda^2} dq \, q \ln \left[ \{q+\alpha(\scl)\}^2 + \beta(\scl) \right]\nn \\
&= \frac{1}{32\pi^2} \left[ I_1 - \alpha(\scl) I_2\right]_{\alpha(\scl)}^{\Lambda^2+\alpha(\scl)} .
\label{(3.15)}
\end{align}

Here, $I_1$ and $I_2$ are given by
\begin{align}
&I_1:= \int dx \, x \ln \left[ x^2 + \beta \right] 
= \frac{1}{2} \left[  \left( x^2 + \beta \right) \ln  \left( x^2 + \beta \right) -x^2 \right], 
\label{(3.16)}
\\
&I_2:= \int dx \ln \left[ x^2 + \beta \right] 
= x \ln  \left( x^2 + \beta \right) -2x  + 2\beta I_A,\label{(3.17)}
\end{align}
and $I_A$ is given by
\begin{align}
I_A:= \int \frac{dx}{  x^2 + \beta }. \label{(3.18)}
\end{align}
The derivations of $I_1$ and $I_2$ are given in Appendix \ref{App}.

\subsection{Calculation of the one-loop potential} 

Our interest is the estimate of the integral in the right-hand side of Eq.~\eqref{(3.15)}.
The integration is performed in Appendix~\ref{App1}, and the result is the following:
\begin{align}
\delta V^{(\phi\ {\rm 1 \mbox{-} loop}) }\left(\scl\right)&=\delta \mathcal{V}^{(\phi\ {\rm 1 \mbox{-} loop}) }\left(\alpha(\scl), \beta(\scl)\right)-\delta \mathcal{V}^{(\phi\ {\rm 1 \mbox{-} loop}) }\left(\alpha(0), \beta(0)\right),\\
\delta \mathcal{V}^{(\phi\ {\rm 1 \mbox{-} loop}) }\left(\alpha, \beta\right)
&:= \frac{1}{32\pi^2}\left[2\alpha \Lambda^2 
+ \frac12 \left( \beta - \alpha^2 \right)
- \frac12 \left( \beta - \alpha^2 \right) \ln \left( \frac{\beta + \alpha^2}{\Lambda^4}\right)
-2 \alpha \beta I_A|_{\alpha}^{\Lambda^2+\alpha}\right].
\label{(3.19)}
\end{align}
Note that $\delta \mathcal{V}^{(\phi\ {\rm 1 \mbox{-} loop}) }\left(\alpha(0), \beta(0)\right)$ is independent of $\scl$.
We can calculate $\beta(\scl)+\alpha^{2}(\scl)$ and $\beta(\scl)-\alpha^{2}(\scl)$ as
\begin{align}
\beta(\scl)+\alpha^{2}(\scl)&=m_1^2 m_2^2 + \lambda_{\sigma\phi}\scl^2,\label{(beta+alpha)}\\
\beta(\scl)-\alpha^{2}(\scl)&=m_1^2 m_2^2 + \lambda_{\sigma\phi}\scl^2 - 2\left(\frac{m_1^2+m_2^2- \xi_{\sigma\phi}\scl^2}{2}\right)^2.\label{(beta-alpha)}
\end{align}
Equation \eqref{(3.19)} includes UV divergent terms,
and they are either absorbed into the cosmological constant (not written explicitly)
or absorbed by the parameters, by imposing the renormalization conditions, as carried out later.
We consider a renormalizable theory where the UV divergences are absorbed by a redefinition of the parameters present in the theory.
The parameters of the original Lagrangian \eqref{(3.3)} are $m_{\sigma}^{2}$, $\lambda_{\sigma}$, $\lambda_{\sigma\phi}$, and $\xi_{\sigma\phi}$.
In Eq.\eqref{(3.19)}, the operators with the UV divergent coefficient $\ln\Lambda^{4}$ are $\sigma_{\rm cl}^{4}$ and $\sigma_{\rm cl}^{2}$,
and the operator with the $\Lambda^{2}$ coefficient is only $\sigma_{\rm cl}^{2}$ [see also Eqs.\eqref{(3.13)} and \eqref{(beta-alpha)}].
Therefore, $\lambda_{\sigma}$ absorbs the $\ln\Lambda^{4}$ divergence, and $m_{\sigma}^{2}$ absorbs both the $\Lambda^{2}$ and the $\ln\Lambda^{4}$ divergences.
In other words, the scalar mass $m_{\sigma}^{2}$ inherently incorporates quadratic and logarithmic divergences and the $\sigma^{4}$ coupling $\lambda_{\sigma}$ incorporates only logarithmic divergence.
As a consequence, these parameters become scale ($N$) dependent.
A study of this effect will be our future work.

Now, our interest is the UV stability. 
UV stability can be understood by analyzing the high-energy behavior of the finite part of Eq.~\eqref{(3.19)}.
The details of the calculation are shown in Appendix A, and the result is given in the next subsection.
Since, for large $\scl$, $\beta$ is always negative, we restrict the analysis in the case with $\beta<0$.
Then, we have
\begin{align}
\beta I_A|_{\alpha}^{\Lambda^2+\alpha}
= 
\frac{1}{2\sqrt{|\beta|}} \left[ \ln \left| \frac{\Lambda^2 + \alpha - \sqrt{|\beta|}}{\Lambda^2 + \alpha + \sqrt{|\beta|}} \right| - \ln \left| \frac{ \alpha - \sqrt{|\beta|}}{ \alpha + \sqrt{|\beta|}} \right| \right] .
\label{(3.20)}
\end{align}
For $\Lambda \to \infty$, $I_A$ turns out to be finite. 
Here $\alpha$ and $\beta$ are given in Eqs.\eqref{(3.13)} and  \eqref{(3.14)}. For small and large $\scl^2$, $\sqrt{|\beta|}$ is obtained explicitly in Appendix \ref{App}. 

\subsection{Properties of the one-loop potential : Stability of the potential}

We evaluate the finite parts of Eq.~\eqref{(3.19)} for large values of $\scl^2$ . The detailed calculation is worked out in Appendix \ref{App5},
and the result is as follows. To extract the finite values from the (UV divergent) effective potential \eqref{(3.5)},
we set the renormalization condition on the coupling constant $\lambda_{\sigma}$, which reads
\begin{align}
\left.\frac{d^{4}V(\scl)}{d\scl^{4}}\right|_{\scl=N}=\lambda_{\sigma}(N)~.
\end{align}
This condition is rather complicated, but to the leading order in $\scl$ , Eq. (\ref{(3.15)}) is reduced to
\begin{align}
\delta V(\scl)^{(\phi\ {\rm 1{\mathchar`-}loop})} = \frac{1}{128\pi^2} \xi_{\sigma\phi}^2 \scl^4 \left(2 \ln\frac{\scl^2}{N^{2}}-1  +\ln \xi_{\sigma\phi}^2  \right)+{\cal O}\left(\scl^2 \right).
\label{(3.24)}
\end{align}

We note that 
\begin{align}
\frac{1}{4}\left(2 \ln\frac{\scl^2}{N^{2}} -1  +\ln \xi_{\sigma\phi}^2  \right) \simeq 
\ln |\scl| + \mbox{const.} \ \to \ \infty 
\qquad \mbox{as} \qquad \scl \to \pm \infty.
\label{(3.25)}
\end{align}

In this paper, we are concerned with the question whether the one-loop corrections due to the HD scalar field (negative norm) differs from that due to the ordinary scalar field (with positive norm), in which respects it differs from the latter.
Hence, we compare the results \eqref{(3.24)} with the known result for the ordinary scalar field, as originally derived by Coleman and Weinberg~\cite{Coleman:1973jx}.  
They found the one-loop correction for a positive-norm scalar field to be\footnote{Note that the normalization of the coupling constant differs from ours, but the precise numerical factor is not essential to our discussion.}
\begin{align}
\frac{\lambda_{\sigma}^2}{256\pi^2} \scl^4 \left(\ln\frac{\scl^2}{N^{2}} + \cdots \right)+\mathcal{O}(\scl^2). 
\label{ns}
\end{align}
Our result, Eq.~\eqref{(3.24)}, for the HD one-loop correction, has essentially the same form as the positive-norm scalar result: the coefficient of the $\scl^4 \ln \scl$ term is positive and proportional to the square of the coupling constant.
It might also be useful to compare our result with the photon one-loop correction in scalar QED (involving the photon $A_\mu$ and a charged scalar field $\sigma$), presented in Sec.~IV of Ref.~\cite{Coleman:1973jx},
\begin{align}
\frac{3e^4}{64\pi^2} \scl^4 \left(\ln\frac{\scl^2}{N^{2}} + \cdots \right)+\mathcal{O}(\scl^2),
\label{photon}
\end{align}
although there is, of course, a fundamental difference between scalar and gauge fields.  
Still, our HD scalar one-loop correction \eqref{(3.24)} shares a similar structure to both Eqs.~\eqref{ns} and \eqref{photon}, particularly regarding the dominant $\scl^4 \ln \scl$ term.

Let us now consider which features of our result, Eq.~\eqref{(3.24)}, are expected and which are nontrivial.
The presence of the $\scl^4 \ln \scl$ structure is anticipated from dimensional analysis.  
Since $\sigma$ has mass dimension 1 [see Eq.~\eqref{(2.6)}], the marginal interaction involving $\sigma$ is $\sigma^4$.  
Because the theory is renormalizable with only marginal interactions, higher-order corrections should also have marginal dimension, i.e., be proportional to $\sigma^4$.  
Moreover, one-loop integrals often yield the logarithmic factor $\ln \sigma$, so the appearance of a $\scl^4 \ln \scl$ term is natural.
However, the positivity of the overall coefficient is nontrivial.  
This is because the total one-loop correction results from the cancellation between positive- and negative-norm contributions arising from the HD scalar propagator.  
The sign of this coefficient is particularly important for determining the UV stability of the potential.
Our calculation demonstrates that the effective potential, including the HD scalar one-loop correction, remains stable as $\scl \to \pm \infty$.  
This is the main result of this paper.

In passing, we evaluate Eq.~\eqref{(3.19)} at small $\scl^2$. 
The full expression of $\delta V(\scl)^{(\phi\ {\rm 1{\mathchar`-}loop})}$ is given in Appendix \ref{App2}; 
it is quite involved, and it makes sense to show it for specific values of $m_1$ and $m_2$ (Appendix \ref{App4}), 
\begin{align}
m_1^2 = m^2, \qquad m_2^2 = 2m^2 . \label{(3.26)}
\end{align}
The result is summarized by writing
\begin{align}
\delta V(\scl)^{(\phi\ {\rm 1{\mathchar`-}loop})} |_{\rm finite} = \frac{1}{32\pi^2} \left[I_1 |_{\alpha}^{\Lambda^2+\alpha} -\alpha I_2 |_{\alpha}^{\Lambda^2+\alpha}\right]_{\rm finite}
= A + B \scl^2 +\mathcal{O}(\scl^4).
\label{(3.27)}
\end{align}
The two coefficients are 
\begin{align}
A &= \frac{m^4}{128 \pi^2}\left[-5+3 \ln2+5 \ln\left(2m^4 \right)\right],  \\
B &= \frac{1}{64\pi^2} \left\{\left[9+4\ln2- 2\ln\left(m^4 \right)\right]\lambda_{\sigma\phi}
+3\left[2+2 \ln2 -\ln\left(m^4 \right) \right]m^2 \xi_{\sigma\phi}\right\}. \label{(3.28)}
\end{align}
The effective potential $V(\scl)$ is given by adding the one-loop term Eq.~\eqref{(3.27)} to the 
tree-level term Eq.~\eqref{(3.4)}:
\begin{align}
V(\scl)^{(0)} +\delta V(\scl)^{(\phi\ {\rm 1{\mathchar`-}loop})} |_{\rm finite} = A + \left(m_{\sigma}^2+ B \right)\scl^2 +{\cal O}\left( \scl^4 \right).
\label{(3.29)}
\end{align}
In fact, by solving the renormalization group (RG) equations, the coupling constants appearing in Eq.~\eqref{(3.27)} acquire a $N$ (renormalization scale) dependence.
The question of the spontaneous symmetry breaking of $U(1)$ can be studied after taking into account this RG effect.
It requires computing many more diagrams, which we wish to leave for a future publication.

\section{CONCLUSIONS}

We have calculated the one-loop effective potential of a complex scalar $\sigma$ in a higher derivative scalar theory, in which a real scalar ﬁeld $\phi$ with a HD kinetic term is coupled to a complex (canonical) scalar $\sigma$ through two types of couplings $\lambda_{\sigma\phi}\sigma^\dagger \sigma \phi^2$ and $\xi_{\sigma\phi}\sigma^\dagger \sigma \left( \partial_\mu \phi^2 \right)$. 
The high-energy behavior of the one-loop effective potential $\delta V^{(\phi\ {\rm 1{\mathchar`-}loop})}(\scl)$ is shown in Eq.~\eqref{(3.24)}. 

We focus on a few important properties of the $\phi$-loop effective potential $\delta V^{(\phi\ {\rm 1{\mathchar`-}loop})}(\scl)$obtained in this paper.
The form of the UV divergent terms ($\Lambda^4$, $\Lambda^2$, $\ln\Lambda$) is as expected from the renormalizable nature of this theory~\cite{Abe:2018rwb}. 
The HD scalar field $\phi$ includes a ghost (negative-norm) component. 
Therefore, it is important to investigate how the finite part of the $\phi$-loop corrections 
differs from the one-loop corrections of an ordinary (positive-norm) field. 
To the best of our knowledge, this has not been previously studied.

Curiously the finite part has turned out to have properties similar to those of the canonical scalar one-loop ~\cite{Coleman:1973jx} in the following sense.
First, the dominant term at high energies is  proportional to $\scl^{4}\ln{\scl}$, a common feature in the (marginally) renormalizable theory.
Second, the coefficient of this term is {\it positive}, which makes the effective potential stable.
%
Within the one-loop contribution from the HD scalar field $\phi$, 
the leading high-energy term is proportional to $\sigma_{\rm cl}^{4}\ln\sigma_{\rm cl}$ with a positive coefficient governed by the derivative coupling $\xi_{\sigma\phi}^{2}$.
In this work, we do not evaluate corrections from the $\sigma$ or gauge loops,
which would generate standard contributions proportional to $\lambda_\sigma^{2}$ and $e^{4}$,
as seen in the original Coleman-Weinberg paper~\cite{Coleman:1973jx}.

A few classes of HD field theories have been studied, HD scalar field theory~\cite{Lee:1969fy}, Lee-Wick's HD gauge theory~\cite{Lee:1970iw}, and quadratic gravity~\cite{Stelle:1976gc}.
Quadratic gravity coupled to the Higgs field shares the same derivative coupling to matter as the HD scalar field theory studied in this paper.
From the particle phenomenology viewpoint, the graviton loop effects at high energies should play a critical role, as mentioned in the Introduction.
However, due to the nonrenormalizable nature of Einstein gravity, uncontrollable divergences appear in the loop evaluation; see Refs.~\cite{Bhattacharjee:2012my, Haba:2014qca, Loebbert:2015eea, Abe:2016irv}.
A naive dimensional analysis  suggests that quadratic gravity may resolve this issue,
and the $\beta$-function of Higgs effective potential at high energies could provide valuable insights into gravity coupled to SM of particle physics.

This way of thinking suggests that a proper understanding of graviton loop effects is essential for exploring the high-energy behavior of the Higgs effective potential.
In this regard, it is also meaningful to study analogous effects in a more tractable setup.
Specifically, analyzing the $\beta$-function in the HD scalar model may provide a useful key.
We leave these topics for the future work.


We have considered here a renormalizable theory of a HD scalar field in $d=4$. 
Theory of a HD scalar field in dimension $d<4$ may be an interesting question. Shinsuke Kawai has brought our attention to the possible extension of the Liouville field theory in this direction and a recent study of a new type of string \cite{Makeenko:2023lef,Makeenko:2021hcm}. 
Yu Nakayama has studied local or global conformal invariance of HD 
theories in $d=2$~\cite{Nakayama:2019xzz}.
He also taught us a study of a statistical mechanical model of such a kind~\cite{Bialek:1987qc}. 

\acknowledgments
We owe very much to Shinsuke Kawai (Sungkyunkwan University), who has made many 
valuable comments at a few stages of this work. 
Y. A. is supported by JSPS Grants-in-Aid for Early-Career Scientists (Grants No. JP19K14719, and No. JP23K13108).
K. I. is supported by Grants-Aid for Scientific Research from Ministry of Education, Science,
Sports and Culture of Japan (Grants No. JP21H05182, No. JP21H05189, and No. JP24K07046).

\appendix

\section{Detailed calculations}
\label{App}

The calculation of the integrals is involved, but it is important to understand the detailed effect. 
This Appendix shows the detailed calculations.

\subsection{$I_1$ and $I_2$ terms}
\label{App1}
The calculation of the one-loop term $\delta V^{(\phi\, {\rm 1{\mathchar`-}loop})}(\sigma_{\rm cl})$ of Eq.~\eqref{(3.15)} is given here. 
The $ I_1$ term is 
\begin{align}
I_1 |_{\alpha}^{\Lambda^2+\alpha} = \frac12  \left\{\left[ \left(\Lambda^2 +\alpha\right)^2 +\beta\right] \ln\left[ \left(\Lambda^2 +\alpha\right)^2 +\beta\right]-\left(\alpha^2 +\beta\right) \ln\left(\alpha^2 +\beta\right) -\left(\Lambda^2 +\alpha\right)^2 +\alpha^2 \right\}.
\label{(A.1.1)}
\end{align}
The right-hand side is easily expanded for large $\Lambda$ and gives
\begin{align}
I_1 |_{\alpha}^{\Lambda^2+\alpha} = \frac12 \left\{ \left(\Lambda^4 +2\alpha\Lambda^2 +\alpha^2 +\beta\right)\ln\Lambda^4 -\Lambda^4 + 3\alpha^2 +\beta-\left(\alpha^2 +\beta\right)\ln\left(\alpha^2 +\beta\right)\right\} + {\cal O}\left( \Lambda^{-2}\right).
\label{(A.1.2)}
\end{align}
The $I_2$ term is 
\begin{align}
I_2 |_{\alpha}^{\Lambda^2+\alpha} =
\left(\Lambda^2 +\alpha\right)\ln\left[\left(\Lambda^2 +\alpha\right)^2 +\beta\right]-\alpha \ln\left(\alpha^2 +\beta\right)-2\left(\Lambda^2 +\alpha\right)+2\alpha+2\beta \left(I_A \bigr|_\alpha^{\Lambda^2+\alpha} \right),
\label{(A.1.3)}
\end{align}
which gives
\begin{align} 
-\alpha I_2 |_{\alpha}^{\Lambda^2+\alpha} =
 -\alpha \Lambda^2 \ln\Lambda^4 -\alpha^2 \ln\Lambda^4 -2\alpha^2 +\alpha^2 \ln\left(\alpha^2 +\beta\right)+2\alpha \Lambda^2-2\alpha\beta\left(I_A\bigr|_{\alpha}^{\Lambda^2+\alpha} \right) + {\cal O}\left( \Lambda^{-2}\right).
\label{(A.1.4)}
\end{align}
The summation of Eqs.~\eqref{(A.1.2)} and \eqref{(A.1.4)} gives
\begin{align}
\left [I_1 -\alpha I_2 \right]_{\alpha}^{\Lambda^2+\alpha}
&= \frac12 \Lambda^4 \ln\Lambda^4 +\frac12 \left(-\alpha^2 +\beta\right) \ln\Lambda^4 -\frac12 \Lambda^4 +2\alpha\Lambda^2 +\frac12\left(-\alpha^2 +\beta\right)
\nn \\
& \hspace{15mm}
+\frac12\left(\alpha^2 -\beta\right)\ln\left(\alpha^2 +\beta\right)-2\alpha\beta\left(I_A \bigr|_{\alpha}^{\Lambda^2+\alpha} \right) + {\cal O}\left( \Lambda^{-2}\right).
\label{(A.1.5)}
\end{align}
The $I_A$ term is shown to be finite below. We have now obtained the finite part as 
\begin{align}
\left.32\pi^2\delta \mathcal{V}^{(\phi\ {\rm 1 \mbox{-} loop}) }\left(\alpha(\scl), \beta(\scl)\right)\right|_{\rm finite}
= \frac12 \left(-\alpha^2 +\beta\right)+\frac12\left(\alpha^2 -\beta\right)\ln\left(\alpha^2 +\beta\right)-2\alpha\beta
\left(I_A|_{\alpha}^{\Lambda^2+\alpha}\right).
\label{(A.1.6)}
\end{align}
The last term in Eq.~\eqref{(A.1.5)} is calculated as follows. $I_A$ is given by
\begin{align}
I_A &=\int \frac{dx}{x^2 +\beta}= \frac{1}{2\sqrt{\beta}} \ln\left|\frac{x-\sqrt{|\beta|}}{x+\sqrt{|\beta|}}\right|  &(\beta<0),\label{(A.1.7a)}
\\
I_A&=\int \frac{dx}{x^2 +\beta}= \frac{1}{\sqrt{\beta}}\tan^{-1} \left(\frac{x}{\sqrt{\beta}}\right)  &(\beta>0).
\label{(A.1.7b)}
\end{align}
For $\beta<0$, Eq.~\eqref{(A.1.7a)} shows
\begin{align}
I_A \bigr|_{\alpha}^{\Lambda^2+\alpha} = \frac{1}{2\sqrt{\beta}} 
\left[ \ln\left|\frac{\Lambda^2+\alpha-\sqrt{|\beta|}}{\Lambda^2+\alpha+\sqrt{|\beta|}}\right| -
 \ln\left|\frac{\alpha-\sqrt{|\beta|}}{\alpha+\sqrt{|\beta|}}\right| 
\right].
\label{(A.1.8)}
\end{align}
The first term in the bracket is $- 2\sqrt{|\beta|}/\Lambda^2\to 0$. 
Hence $I_A$ is finite.
\begin{align}
\lim_{\Lambda \to \infty} I_A \bigr|_{\alpha}^{\Lambda^2+\alpha} = -\frac{1}{2\sqrt{\beta}}
 \ln\left|\frac{\alpha-\sqrt{|\beta|}}{\alpha+\sqrt{|\beta|}}\right|.
\label{(A.1.9)}
\end{align}
For $\beta>0$, Eq.~\eqref{(A.1.7b)} should be taken,
\begin{align}
I_A \bigr|_{\alpha}^{\Lambda^2+\alpha} =\frac{1}{\sqrt{\beta}}\left[\tan^{-1} \left(\frac{\Lambda^2+\alpha}{\sqrt{\beta}}\right)-\tan^{-1} \left(\frac{\alpha}{\sqrt{\beta}}\right) \right].
\label{(A.1.10)}
\end{align}
The first term in the bracket is $\tan^{-1} (\infty)=\pi/2$. 
Hence, 
\begin{align}
\lim_{\Lambda \to \infty} I_A \bigr|_{\alpha}^{\Lambda^2+\alpha} =-\frac{1}{\sqrt{\beta}}\tan^{-1} \left(\frac{\alpha}{\sqrt{\beta}}\right),
\label{(A.1.11)}
\end{align}
where we have eliminated the constant term $\pi/2$, which can be absorbed into the cosmological constant.

Although Eqs. \eqref{(A.1.9)} and \eqref{(A.1.11)} can be expressed in terms of $\sigma_\text{cl}^2$, the resulting expressions are rather complicated.
Therefore, we instead discuss the approximate expression for the large-$\sigma_\text{cl}^2$ and small-$\sigma_\text{cl}^2$ region in the following subsections of Appendix A.

\subsection{Large-$\scl^2$ behavior} 
\label{App5}

We study, here, the large-$\scl^2$ behavior of the one-loop corrections to $V(\scl)$. 
To this end, we assume 
\begin{align}
\lambda_{\sigma\phi},\ m_1^2 +m_2^2 \ll \scl^2, \label{(A.5.1)}
\end{align}
and set 
\begin{align}
&\alpha= \frac{ m_1^2 +m_2^2-\xi_{\sigma\phi}\scl^2}{2} \qquad (\mbox{$\alpha<0$ if $\xi_{\sigma\phi}>0$}) ,
\label{(A.5.2)} \\
&\beta\simeq  \frac{2\left[2\lambda_{\sigma\phi}+\left(m_1^2 +m_2^2\right)\xi_{\sigma\phi}\right]\scl^2-\xi_{\sigma\phi}^2\scl^4}{4} <0.
\label{(A.5.3)}
\end{align}
We need 
\begin{align}
&\sqrt{|\beta|}
\simeq \pm\left[\frac{\xi_{\sigma\phi}\scl^2  -\left(m_1^2 +m_2^2\right)}{2} -\frac{\lambda_{\sigma\phi}}{\xi_{\sigma\phi}}\right]
&\hspace{-25mm} \left(\mbox{ $+$ for $\xi_{\sigma\phi}>0$ or $-$ for $\xi_{\sigma\phi}<0$}\right), \label{(A.5.4)}\\
&\alpha\pm\sqrt{|\beta|}
\simeq -\frac{\lambda_{\sigma\phi}}{\xi_{\sigma\phi}} 
&\hspace{-25mm} \left(\mbox{ $+$ for $\xi_{\sigma\phi}>0$ or $-$ for $\xi_{\sigma\phi}<0$}\right),
\\
&\alpha\mp\sqrt{|\beta|}
\simeq -\xi_{\sigma\phi}\scl^2 + m_1^2 +m_2^2  +\frac{\lambda_{\sigma\phi}}{\xi_{\sigma\phi}} 
&\hspace{-25mm} \left(\mbox{ $-$ for $\xi_{\sigma\phi}>0$ or $+$ for $\xi_{\sigma\phi}<0$}\right),\label{(A.5.5)}\\
&\ln \left|\frac{\alpha+\sqrt{|\beta|}}{\alpha-\sqrt{|\beta|}}\right|
\simeq \mp \left\{ \ln \frac{\scl^2}{\left|\lambda_{\sigma\phi} \right|} +\ln \xi_{\sigma\phi}^2  \right\}
&\hspace{-25mm} \left(\mbox{ $-$ for $\xi_{\sigma\phi}>0$ or $+$ for $\xi_{\sigma\phi}<0$}\right) ,\label{(A.5.9)}\\
&\alpha^2 +\beta\simeq \lambda_{\sigma\phi}\scl^2, \\
&-\alpha^2 +\beta\simeq -\frac{\xi_{\sigma\phi}^2 \scl^4}{2} +\left[\lambda_{\sigma\phi}+\left(m_1^2 +m_2^2 \right)\xi_{\sigma\phi}\right]\scl^2, \label{(A.5.6)} \\
&\alpha\sqrt{|\beta|}
\simeq\pm \left\{ -\frac{\xi_{\sigma\phi}^2 \scl^4}{4} +\frac{\lambda_{\sigma\phi}+\left(m_1^2 +m_2^2 \right)\xi_{\sigma\phi}}{2}\scl^2 \right\} 
\nn \\
&&\hspace{-25mm} 
\left(\mbox{ $+$ for $\xi_{\sigma\phi}>0$ or $-$ for $\xi_{\sigma\phi}<0$}\right). 
 \label{(A.5.7)}
\end{align}
It is relatively easy to calculate the finite part \eqref{(A.1.9)}. 
We recall 
\begin{align}
-2\alpha \beta  I_A \bigr|_{\alpha}^{\Lambda^2+\alpha}
&\simeq \alpha \sqrt{|\beta|} \ln \left|\frac{\alpha+\sqrt{|\beta|}}{\alpha-\sqrt{|\beta|}}\right|
\nn \\
&=  \left\{ \ln \frac{\scl^2}{\left|\lambda_{\sigma\phi}\right|}  +\ln \xi_{\sigma\phi}^2  \right\}\frac{\xi_{\sigma\phi}^2}{4} \scl^4 +{\cal O}\left(\scl^2 \right) .
\label{(A.5.10)}
\end{align}
The remaining two terms are 
\begin{align}
&\frac12\left(-\alpha^2 +\beta\right)+\frac12\left(\alpha^2 -\beta\right)\ln\left(\alpha^2 +\beta\right)
= \left[-1+\ln\left(\left|\lambda_{\sigma\phi} \right|\scl^2\right)\right]\frac{\xi_{\sigma\phi}^2 \scl^4}{4}+{\cal O}\left(\scl^2 \right).
\label{(A.5.11)}
\end{align}
Adding Eqs.~\eqref{(A.5.11)} and \eqref{(A.5.10)}, 
\begin{align}
\left.\left[I_1-\alpha I_2\right]_{\alpha}^{\Lambda^2+\alpha} \right|_{\rm finite}
= \frac14 \xi_{\sigma\phi}^2 \scl^4\left(2 \ln\scl^2-1  +\ln \xi_{\sigma\phi}^2  \right)+{\cal O}\left(\scl^2 \right).
\label{(A.5.13)}
\end{align}
Summarizing the above result, the dominant behavior at large $\scl$ is given by  
\begin{align}
\delta V(\scl)^{(\phi\ {\rm 1{\mathchar`-}loop})} = \frac{1}{128\pi^2} \xi_{\sigma\phi}^2 \scl^4\left(2 \ln\scl^2-1  +\ln \xi_{\sigma\phi}^2  \right)+{\cal O}\left(\scl^2 \right).
\label{(A.5.15)}
\end{align}

\subsection{Small-$\scl^2$ behavior} 
\label{App2}
We first study the small-$\scl^2$ behavior, ignoring terms of ${\cal O}\left( \sigma_{\rm cl}^{4} \right)$.
Here, we assume 
\begin{align}
m_2^2 > m_1^2 > 0.
\label{(A.1.12)}
\end{align}

In the small-$\scl^2$ approximation, we have
\begin{align}
&
\alpha= \frac{m_1^2 +m_2^2 -\xi_{\sigma\phi} \scl^2}{2}, \label{(A.2.1)}
\\
&
\beta\simeq -\frac{\left(m_2^2 -m_1^2\right)^2}{4} +\left[\lambda_{\sigma\phi}+\frac{m_1^2 +m_2^2 }{2}\xi_{\sigma\phi}\right]\scl^2 < 0, \label{(A.2.2)}\\
&
\sqrt{|\beta|}
\simeq 
 \frac{m_2^2 -m_1^2}{2} -\frac{\left[2\lambda_{\sigma\phi}+\left(m_1^2 +m_2^2\right)\xi_{\sigma\phi}\right]\scl^2}{2\left(m_2^2 -m_1^2 \right)}, \label{(A.2.3)}\\
&\ln\left|\frac{\alpha+\sqrt{|\beta|}}{\alpha-\sqrt{|\beta|}}\right|
\simeq
\ln\left\{\left[m_2^2 -\frac{\left(\lambda_{\sigma\phi}+m_2^2 \xi_{\sigma\phi}\right)\scl^2 }{m_2^2 -m_1^2} \right]\frac{m_2^2 -m_1^2}{m_1^2 \left(m_2^2 -m_1^2 \right) +\left(\lambda_{\sigma\phi}+m_1^2 \xi_{\sigma\phi}\right)\scl^2} \right\}
\nn \\ & \hspace{24mm}
\simeq
 \ln\left(\frac{m_2^2}{m_1^2} \right)-\left[\frac{m_1^2 +m_2^2}{m_1^2 m_2^2}\lambda_{\sigma\phi} +2\xi_{\sigma\phi}\right]\frac{\scl^2}{m_2^2 -m_1^2},
\label{(A.4.3)}\\
&
\alpha+\sqrt{|\beta|}\simeq
m_2^2 -\frac{\left(\lambda_{\sigma\phi}+m_2^2 \xi\right)\scl^2}{m_2^2 -m_1^2 }, \\
&
\alpha-\sqrt{|\beta|}\simeq
m_1^2 +\frac{\left(\lambda_{\sigma\phi}+m_1^2 \xi\right)\scl^2}{m_2^2 -m_1^2 }, \label{(A.2.4)}\\
&
\alpha^2+\beta
~=~
 m_1^2 m_2^2 +\lambda_{\sigma\phi} \scl^2,\\
& -\alpha^2+\beta
~=~ -\frac{m_1^4 +m_2^4 }{2} +\left[\lambda_{\sigma\phi}+\left(m_1^2 +m_2^2 \right)\xi_{\sigma\phi}\right]\scl^2 , \label{(A.2.5)}\\
&
\alpha\sqrt{|\beta|}
\simeq
 \frac{m_2^4 -m_1^4 }{4} -\frac{\left[\left(m_1^2 +m_2^2\right)\lambda_{\sigma\phi}+\left(m_1^4 +m_2^4 \right)\xi_{\sigma\phi}\right]\scl^2}{2\left(m_2^2 -m_1^2 \right)}.
 \label{(A.2.6)}
\end{align}
The quantities look rather complicated. 



We first evaluate the $I_A$ term of Eq.~\eqref{(A.1.9)}, 
\begin{align}
-2\alpha\beta \left ( I_A \bigr|_{\alpha}^{\Lambda^2+\alpha} \right)
&
=- \alpha\sqrt{|\beta|} 
\ln\left|\frac{\alpha+\sqrt{|\beta|}}{\alpha-\sqrt{|\beta|}}\right| \nn \\
&\simeq \frac{m_2^4-m_1^4}{4}\ln\left(\frac{m_2^2}{m_1^2} \right)
-\left\{\left[\frac{\left(m_1^2 +m_2^2 \right)^2}{m_1^2 m_2^2} +2\frac{m_1^2 +m_2^2}{m_2^2 -m_1^2}\ln\left(\frac{m_2^2}{m_1^2} \right)\right]\frac{\lambda_{\sigma_\phi}}{4}
\right. \nn \\ & \left.
-\left[\left(m_1^2 +m_2^2\right)+ \frac{m_1^4 +m_2^4}{m_2^2 -m_1^2}\ln \left(\frac{m_2^2}{m_1^2} \right)\right]\frac{\xi_{\sigma\phi}}{2}\right\}\scl^2.
\label{(A.4.4)}
\end{align}
The remaining two terms of Eq.~\eqref{(A.1.9)} are easier to calculate. 
\begin{align}
\frac12 & \left(-\alpha^2 +\beta\right)+\frac12 \left(\alpha^2 -\beta\right)\ln\left(\alpha^2 +\beta\right)
\nn \\ &
\simeq -\left[1-\ln\left(m_1^2 m_2^2 \right)\right]\frac{m_1^4 +m_2^4}{4}
\nn \\ & \qquad
-\left\{\frac{m_1^4 +m_2^4}{2m_1^2 m_2^2}\lambda_{\sigma\phi} + \left[1-\ln\left(m_1^2 m_2^2 \right)\right]\left[\lambda_{\sigma\phi}+\left(m_1^2 +m_2^2\right) \xi_{\sigma\phi}\right]\right\}\frac{\scl^2}{2}.
\label{(A.4.5)}
\end{align}
Adding Eqs.~\eqref{(A.4.4)} and \eqref{(A.4.5)} we obtain 
\begin{align}
\left.\left[I_1-\alpha I_2\right]_{\alpha}^{\Lambda^2+\alpha} \right|_{\rm finite}
\simeq 
&\frac{m_2^4 -m_1^4}{4}\ln\left(\frac{m_2^2}{m_1^2} \right)
-\left[1-\ln\left(m_1^2 m_2^2 \right)\right]\frac{m_1^4 +m_2^4}{4}
\nn \\ &
-\left\{\left[\frac{\left(m_1^2 +m_2^2 \right)^2}{m_1^2 m_2^2}
+\frac{m_1^2 +m_2^2}{m_2^2 -m_1^2}\ln\left(\frac{m_2^2}{m_1^2}\right)
-\ln\left(m_1^2 m_2^2\right) \right]\frac{\lambda_{\sigma\phi}}{2}
\right. \nn \\ &\left.
+\left[2\left(m_1^2 +m_2^2 \right)+\frac{m_1^4 +m_2^4}{m_2^2 -m_1^2}\ln \left(\frac{m_2^2}{m_1^2}\right) - \left(m_1^2+m_2^2 \right)\ln \left(m_1^2m_2^2 \right)\right]\frac{\xi_{\sigma\phi}}{2}\right\}\scl^2.
\label{(A.4.6)}
\end{align}
The above expression is quite complicated. It would be useful to consider cases with specific values of $m_1$ and $m_2$. 
We consider two cases; one is a case with $m_1^2 = 0$ and $m_2^2 =m^2 > 0$, and the other is with $m_1^2 = m^2$ and $m_2^2= 2m^2$.

\subsubsection{Case with $m_1^2 = 0$ and $m_2^2 =m^2 > 0$}

We choose 
\begin{align}
m_1^2 = 0  \qquad \left(m_2^2 =m^2 > 0\right), \label{(A.3.1)}
\end{align}
and then we have
\begin{align}
&\alpha=\frac{m^2 - \xi_{\sigma\phi}\scl^2}{2}, \label{(A.3.2)}
\\
&\beta\simeq \frac{-m^4 +\left(4\lambda_{\sigma\phi}+2m^2 \xi_{\sigma\phi} \right)\scl^2}{4},
\label{(A.3.3)}\\
&\sqrt{|\beta|} \simeq \frac{m^2}{2} -\left(\frac{\lambda_{\sigma\phi}}{m^2}+\frac{\xi_{\sigma\phi}}{2} \right)\scl^2, \label{(A.3.4)}
\\
&\alpha+\sqrt{|\beta|}\simeq m^2 -\left(\frac{\lambda_{\sigma\phi}}{m^2} +\xi_{\sigma\phi}\right)\scl^2,
\\
&\alpha-\sqrt{|\beta|}\simeq \frac{\lambda_{\sigma\phi}}{m^2}\scl^2 , \label{(A.3.5)}
\\
&\ln \left|\frac{\alpha+\sqrt{|\beta|}}{\alpha-\sqrt{|\beta|}} \right|
\simeq \ln\left|\left[m^2 -\left(\frac{\lambda_{\sigma\phi}}{m^2} +\xi_{\sigma\phi}\right)\scl^2 \right]\frac{m^2}{\lambda_{\sigma\phi}\scl^2} \right|
\nn \\ & \hspace{24mm}
\simeq \ln\left(\frac{m^4}{\left|\lambda_{\sigma\phi} \right|\scl^2} \right)-\left(\frac{\lambda_{\sigma\phi}}{m^4} +\frac{\xi_{\sigma\phi}}{m^2}\right)\scl^2 ,
\label{(A.3.6)}
\\
&\alpha^2 +\beta= \lambda_{\sigma\phi} \scl^2,
\\
&-\alpha^2 +\beta= -\frac{m^4+2\left(\lambda_{\sigma\phi}+m^2 \xi_{\sigma\phi} \right)\scl^2}{2} , \label{(A.3.7)}
\\
&\alpha\sqrt{|\beta} \simeq 
\frac{ m^4 }{4}\left[1-2\left(\frac{\lambda_{\sigma\phi}}{m^4} +\frac{\xi_{\sigma\phi}}{m^2}\right)\scl^2 \right].
\label{(A.3.8)}
\end{align}

We are now ready to compute Eq.~\eqref{(A.1.9)} (for $\beta<0$):
\begin{align}
-2\alpha\beta \left (I_A |_{\alpha}^{\Lambda^2+\alpha} \right)
&
= \alpha\sqrt{|\beta|} 
\ln\left|\frac{\alpha+\sqrt{|\beta|}}{\alpha-\sqrt{|\beta|}}\right|
\nn \\ 
&
= \frac{m^4}{4} \left[1-2\left(\frac{\lambda_{\sigma\phi}}{m^4} +\frac{\xi_{\sigma\phi}}{m^2}\right)\scl^2 \right] 
\left[\ln\left(\frac{m^4}{\left|\lambda_{\sigma\phi} \right|\scl^2}\right)-\left(\frac{\lambda_{\sigma\phi}}{m^4} +\frac{\xi_{\sigma\phi}}{m^2}\right)\scl^2  \right].
\label{(A.3.9)}
\end{align}
Near $\scl=0$, 
\begin{align}
-2\alpha\beta \left (I_A |_{\alpha}^{\Lambda^2+\alpha} \right)= \frac{m^4}{4} \ln\left(\frac{m^4}{\left|\lambda_{\sigma\phi} \right|\scl^2}\right).
\label{(A.3.10)}
\end{align}
Curiously Eq.~\eqref{(A.3.6)} is singular at $\scl=0$.
The reason for this singularity is not immediately clear; it may be due to the fact that there is a massless particle, $m_1^2 =0$.

\subsubsection{Case with $m_1^2 = m^2$ and $m_2^2= 2m^2$}
\label{App4}

For calculational ease we may study other values of $m_1^2$ and $m_2^2$ , but avoiding the singularity at $\scl=0$, 
\begin{align}
m_1^2 = m^2, \qquad m_2^2= 2m^2 .\label{(A.3.12)}
\end{align}
Then 
\begin{align}
&\alpha= \frac{3m^2-\xi_{\sigma\phi}\scl^2}{2},\label{(A.3.13)}
\\
&\beta\simeq \frac{-m^4 +\left(4\lambda_{\sigma\phi}+6m^2 \xi_{\sigma\phi} \right)\scl^2}{4},
\label{(A.3.14)}
\\
&\sqrt{|\beta|} \simeq \frac{m^2}{2} -\left(\frac{\lambda_{\sigma\phi}}{m^2}+\frac{3\xi_{\sigma\phi}}{2} \right)\scl^2.
\label{(A.3.15)}
\\
&\alpha+\sqrt{|\beta|}\simeq 2 m^2 -\left(\frac{\lambda_{\sigma\phi}}{m^2} +2 \xi_{\sigma\phi}\right)\scl^2,
\\
&\alpha-\sqrt{|\beta|}\simeq  m^2 +\left(\frac{\lambda_{\sigma\phi}}{m^2} + \xi_{\sigma\phi}\right)\scl^2,
\label{(A.3.16)}
\\
&\ln \left|\frac{\alpha+\sqrt{|\beta|}}{\alpha-\sqrt{|\beta|}} \right|
\simeq \ln2 - \left(\frac{3\lambda_{\sigma\phi}}{2m^4} +\frac{2\xi_{\sigma\phi}}{m^2}\right)\scl^2 , \label{(A.3.17)}
\\
&\alpha^2 +\beta\simeq 2m^4+ \lambda_{\sigma\phi} \scl^2,
\\
&-\alpha^2 +\beta\simeq \frac{-5m^4+2\left(\lambda_{\sigma\phi}+3m^2 \xi_{\sigma\phi} \right)\scl^2}{2} ,
\label{(A.3.18)}
\\
&\alpha\sqrt{|\beta|} 
\simeq \frac{ 3 m^4 }{4}\left[1-2\left(\frac{\lambda_{\sigma\phi}}{m^4} +\frac{5\xi_{\sigma\phi}}{3m^2}\right)\scl^2 \right].
\label{(A.3.19)}
\end{align}
We first evaluate the last term $I_A$ (for $\beta<0$). 
Substitute Eqs.~\eqref{(A.3.17)} and \eqref{(A.3.19)} into $\alpha \beta I_A$.
\begin{align}
-2\alpha\beta \left (I_A |_{\alpha}^{\Lambda^2+\alpha} \right)
&
= \alpha\sqrt{|\beta|} 
\ln\left|\frac{\alpha+\sqrt{|\beta|}}{\alpha-\sqrt{|\beta|}}\right| \nn \\
&
= \frac{m^4}{4} \left[1-2\left(\frac{\lambda_{\sigma\phi}}{m^4} +\frac{\xi_{\sigma\phi}}{m^2}\right)\scl^2 \right] 
\left[\ln\left(\frac{m^4}{\lambda_{\sigma\phi}\scl^2}\right)-\left(\frac{\lambda_{\sigma\phi}}{m^4} +\frac{\xi_{\sigma\phi}}{m^2}\right)\scl^2  \right].
\label{(A.3.20)}
\end{align}
   The first two terms in Eq.~\eqref{(A.1.6)} are easier to calculate. Using Eq.~\eqref{(A.3.18)}, 
\begin{align}
\frac12 & \left(-\alpha^2 +\beta\right)+\frac12 \left(\alpha^2 -\beta\right)\ln \left(\alpha^2 +\beta \right) 
\nn \\ &
\simeq 
\frac54 m^4 \left[1- \ln\left(2m^4 \right)\right]+\left\{\left[\frac94 -\ln\left(2m^4 \right)\right]\lambda_{\sigma\phi}-3\left[1-\ln \left(2m^4 \right) \right]m^2 \xi_{\sigma\phi}\right\}\scl^2 . 
\label{(A.3.21)}
\end{align}
Adding the three terms,  
\begin{align}
\left.\left[I_1-\alpha I_2\right]_{\alpha}^{\Lambda^2+\alpha} \right|_{\rm finite}
\simeq 
& \left[-\frac54+\frac34 \ln2+\frac54 \ln\left(2m^4 \right)\right]m^4 
\nn \\ &
+\left\{\left[\frac92+2\ln2- \ln\left(m^4 \right)\right]\lambda_{\sigma\phi}
+\left[3+3 \ln2 -\frac32\ln\left(m^4 \right) \right]m^2 \xi_{\sigma\phi}\right\}\scl^2 .
\label{(A.3.22)}
\end{align}

\section{Dimensional Regularization}
\label{App of Dimensional Regularization}
The momentum integral in Eq.~\eqref{(3.9)} is UV divergent and has to be regularized.
In this paper, we have computed in the momentum cutoff method.
Other regularizations are known, but they are known to give the same result as far as the finite quantities are concerned. 

We rewrite Eq.~\eqref{(3.9)} as
\begin{align}
\delta V^{(\phi\ {\rm 1{\mathchar`-}loop}) }\left(\scl\right)
&=\frac12 \int  \frac{d^4 k_E}{(2\pi)^4} \ln \left[{\tilde {\cal O}}_{\tilde \phi} (k_E) \right]\nonumber\\
&=\frac12 \int  \frac{d^4 k_E}{(2\pi)^4} \left[\ln \left(k_E^{2}+\alpha+\sqrt{-\beta}\right)+\ln \left(k_E^{2}+\alpha-\sqrt{-\beta}\right)\right],
\label{(B.1)}
\end{align}
where
\begin{align}
\alpha+\sqrt{-\beta}=\frac{m_1^2+m_2^2- \xi_{\sigma\phi}\scl^2}{2}+\sqrt{-m_1^2 m_2^2 - \lambda_{\sigma\phi}\scl^2 + \left(\frac{m_1^2+m_2^2- \xi_{\sigma\phi}\scl^2}{2}\right)^2},
\label{(B.2)}\\
\alpha-\sqrt{-\beta}=\frac{m_1^2+m_2^2- \xi_{\sigma\phi}\scl^2}{2}-\sqrt{-m_1^2 m_2^2 - \lambda_{\sigma\phi}\scl^2 + \left(\frac{m_1^2+m_2^2- \xi_{\sigma\phi}\scl^2}{2}\right)^2}.
\label{(B.3)}
\end{align}

In dimensional regularization, the $d$-dimensional integral is given by:
\begin{align}
\int \frac{d^d k_E}{(2\pi)^d} \ln \left(k_E^{2}+\bigtriangleup\right)&=-\frac{1}{\left(4\pi\right)^{\frac{d}{2}}}\frac{1}{\frac{d}{2}\left(\frac{d}{2}-1\right)}\Gamma\left(2-\frac{d}{2}\right)\bigtriangleup^{\frac{d}{2}}\nonumber\\
&=-\frac{1}{32\pi^{2}}\bigtriangleup^{2}\left(\frac{2}{\epsilon}+\frac{3}{2}-\gamma-\ln\bigtriangleup+\ln4\pi\right)+\mathcal{O}(\epsilon).
\label{(B.4)}
\end{align}
When applying dimensional regularization to a four-dimensional theory, we set $d=4-\epsilon$, as usual.

The application of Eq.~\eqref{(B.4)} to Eq.~\eqref{(B.1)} gives
\begin{align}
\delta V^{(\phi\ {\rm 1{\mathchar`-}loop}) }\left(\scl\right)
&=\frac{1}{64\pi^{2}}\Biggl[2\left(\beta-\alpha^{2}\right)\cdot\frac{2}{\epsilon}+2\left(\beta-\alpha^{2}\right)\left(\frac{3}{2}-\gamma+\ln4\pi\right)\nonumber\\
&\qquad\qquad\quad-\left(\beta-\alpha^{2}\right)\ln{\left(\beta+\alpha^{2}\right)}-2\alpha\sqrt{-\beta}\ln{\frac{\alpha-\sqrt{-\beta}}{\alpha+\sqrt{-\beta}}}\Biggr].
\label{(B.11)}
\end{align}
Comparing Eq.~\eqref{(3.19)} and Eq.~\eqref{(B.11)}, 
we can note that the divergent terms and the finite terms correspond to
\begin{align}
&\frac{1}{64\pi^{2}}\left(\beta-\alpha^{2}\right) \ln \Lambda^{4}=\frac{1}{64\pi^{2}}\cdot2\left(\beta-2\alpha^{2}\right)\ln \Lambda^{2}\quad\Leftrightarrow\quad\frac{1}{64\pi^{2}}\cdot2\left(\beta-2\alpha^{2}\right)\cdot\frac{2}{\epsilon},
\label{(B.12)}\\
&\frac{1}{64\pi^{2}}\left[\left(\beta-\alpha^{2}\right)-\left(\beta-\alpha^{2}\right)\ln{\left(\beta+\alpha^{2}\right)} - 4\alpha\beta I_A|_{\alpha}^{\Lambda^2+\alpha}\right]\nonumber\\
\quad\Leftrightarrow\quad&
\frac{1}{64\pi^{2}}\left[2\left(\beta-\alpha^{2}\right)\left(\frac{3}{2}-\gamma+\ln4\pi\right)-\left(\beta-\alpha^{2}\right)\ln{\left(\beta+\alpha^{2}\right)}-2\alpha\sqrt{-\beta}\ln{\frac{\alpha-\sqrt{-\beta}}{\alpha+\sqrt{-\beta}}}\right].
\label{(B.13)}
\end{align}
We note that the power divergent terms $\cfrac{1}{64\pi^{2}}\left[\Lambda^{4} \ln \Lambda^{4}  - \Lambda^{4} + 4\alpha  \Lambda^{2}\right]$,
which are present in the cutoff method [see Eq.~\eqref{(A.1.5)}], are absent in the dimensional method.

\bibliography{ref}

\end{document}